\documentstyle[aps]{revtex}
\input{epsf.tex}
\begin{document}
\draft
\title{Feshbach resonance in dense ultracold Fermi gases}
\author{R. Combescot}
\address{Laboratoire de Physique Statistique,
 Ecole Normale Sup\'erieure*,
24 rue Lhomond, 75231 Paris Cedex 05, France}
\date{Received \today}
\maketitle

\begin{abstract}
We propose a coherent framework allowing to deal with many-body 
effects in dense ultracold Fermi gases in the presence of a Feshbach 
resonance. We show that the simple effect of Pauli exclusion induces a 
strong modification of the basic scattering properties, leading in 
particular to an energy dependence of the effective scattering length on 
the scale of the chemical potential. This results in a smearing of the 
Feshbach resonance and provides a natural explanation for recent 
experimental findings.
\end{abstract}

Besides the continuing progress in understanding Bose Einstein 
condensation in ultracold bosonic atomic gases, the exploration of 
similar dense \cite{jin} fermionic systems has received recently a strong 
impetus from experiments reaching the strongly degenerate regime with 
mixtures of fermions in two different hyperfine states \cite{thomas}. A 
main purpose in exploring these systems is the search for a transition to 
a BCS superfluid \cite{stoofal}. A particularly fascinating possibility of 
experimentally controling the strength of the interaction, already 
demonstrated in Bose systems, has been used in these last experiments. 
It consists in working in the vicinity of a Feshbach resonance 
\cite{timm} where the scattering length, directly linked to the 
interaction, depends strongly on the applied magnetic field. It has been 
emphasized recently \cite{holland} that the strong interaction obtained 
in the vicinity of this resonance, together with the Bose condensation of 
the molecules corresponding to the underlying bound state, could be of 
major interest for the BCS transition. 

In the present paper we set up a theoretical framework for handling 
many-body effects in the presence of a Feshbach resonance. As a first 
consequence we show that the simple effect of Pauli exclusion explains 
qualitatively that the Feshbach resonance is strongly smeared for 
degenerate gases, as it appears experimentally \cite{thomas}. Here we 
will restrict ourselves for simplicity to the case where the (quasi) bound 
state responsible for the resonance is not thermally occupied, which 
corresponds to a negative effective scattering length. Similarly we will 
consider only the normal state although our formalism can be extended 
to the superfluid state.

First we model the Feshbach resonance in the following way. The 
Feshbach resonance \cite{timm} is actually produced by the somewhat 
complex interplay between the spin (electronic and nuclear) and orbital 
degrees of freedom. However in order to have an understanding of the 
physics linked to this resonance, it is convenient to use a simple 
modeling where the spin degrees of freedom are no longer involved. 
One can think that the resonance occurs because there is a (quasi) bound 
state caused by the existence of a deep well in the atomic interaction 
potential at short distances. One will indeed have a bound state, 
corresponding physically to the formation of a molecule, if this well is 
essentially isolated from the large distance region by a high barrier with 
very small transmission probability. Clearly this model behaves 
phenomenologically exactly in the same way as a Feshbach resonance 
\cite{landau} .

On the other hand we are only interested in the effect of this molecular 
state on the scattering of two atoms, and we will ignore the effect of the 
other atoms on these two when they are close together. This is 
reasonable since these other atoms will be most of the time far away and 
will not perturbe the two atoms we consider. In other words we make 
use of the fact that the gas is dilute on the scale of the molecular state. 
Naturally this means that we ignore for example the possibility that 
three atoms are close together, which would require additional 
ingredients in the description.

We further simplify our description by separating the possible distance 
between the two particles in two domains. Either they are far away and 
interact by the long range part of the potential (this range will be 
denoted by index 1). Or they are quite close and interact through the 
deep well of the potential (this is noted by index 2). More precisely one 
can define the boundary between the two domains as being at some 
distance R, large compared to the molecular size but small with respect 
to interparticle distance. Then instead of describing exactly the transition 
from domain 1 to domain 2, we assume that a term in the Hamiltonian 
gives rise to a matrix element producing this transition. We note that 
this problem is actually quite similar to the one raised by tunneling of 
electrons between two metallic electrodes through an insulating barrier 
and our approach is at the level of the tunneling Hamiltonian 
\cite{bardeen} . It is actually known that this simple modeling can be 
improved up to the exact problem, and that many-body effects can also 
be included in this theory \cite{caroli}. 
However this does not seem necessary in the 
present case and our simple modeling should be a quite enough.

To be more specific now consider first the problem with the center of 
mass of the two atoms at rest, so we deal with a one body problem with 
interparticle separation $ {\bf  r}$ and reduced mass $ m _{r} = m/2$. 
The parts of the Hamiltonian $H$ corresponding respectively to domain 
1 and 2 are:
\begin{eqnarray}
H _{11}= \frac{ {\bf  p} ^{2}}{2 m _{r}} + V _{1}( {\bf  r})  
\hspace{2cm}
H _{22}= \frac{ {\bf  p} ^{2}}{2 m _{r}} + V _{2}( {\bf  r})
\label{eq1}
\end{eqnarray}
where $ V _{1}( {\bf  r}) $ and $ V _{2}( {\bf  r}) $ are the interaction 
potential at long and short distance respectively. For example $ V _{1}( 
{\bf  r}) $ gives rise to a scattering length $a _{bg} $ while $ V _{2}( {\bf  r}) 
$ produces the molecular state. Then we describe hopping between the 
two domains by the non-local term:
\begin{eqnarray}
H _{12}( {\bf  r}, {\bf  r}') = t ( {\bf  r}, {\bf  r}')
\label{eq2}
\end{eqnarray}
with $ H _{21}= H ^{\dagger}_{12}$. Since the purpose of $ t ( {\bf  
r}, {\bf  r}')$ is just to make the particle cross the boundary between 
domain 1 and 2, we can assume it to be short-ranged around this 
boundary.

Let now $ G _{11} ^{0}( {\bf  r}, {\bf  r}', \omega )$ be the 
propagator corresponding to $ H _{11} $, at frequency $ \omega $ (we 
take $ \hbar = 1$) and similarly $ G _{22} ^{0}( {\bf  r}, {\bf  r}', 
\omega )$ the propagator corresponding to $ H _{22} $. Physically 
they describe the motion either at long or short distance, without the 
possibility to hop between the two domains. We treat now exactly the 
effect of the hopping term. We introduce the full propagator $ G ( {\bf  
r}, {\bf  r}', \omega )$ corresponding to $H$. If ${\bf  r}$ and ${\bf  
r}'$ are large (we call again $ G _{11}$ the corresponding 
propagator), the particle can propagate either by staying in domain 1 
(this is described by $ G _{11} ^{0}$) or, after a stay in domain 1, by 
making a first hop to domain 2 and then propagating by any means back 
to domain 1. This leads to the following equation between operators:
\begin{eqnarray}
G _{11}= G _{11} ^{0} + G _{11} ^{0} t G _{21}
\label{eq3}
\end{eqnarray}
where a same frequency is understood in all the propagators. We have a 
similar relation for propagation between domain 2 and 1, except that 
hoping is now required (because $ G _{21} ^{0} = 0 $):
\begin{eqnarray}
G _{21}= G _{22} ^{0} t^{\dagger} G _{11}
\label{eq4}
\end{eqnarray}
 Carrying Eq.(4) into Eq.(3) gives:
\begin{eqnarray}
G _{11}= G _{11} ^{0} + G _{11} ^{0} t G _{22} ^{0} t^{\dagger} G _{11}
\label{eq5}
\end{eqnarray}
which is an integral equation for $ G _{11}$.

This can now be simplified if we take into account that a single bound 
state of $ H _{22}$, corresponding to the Feshbach resonance we are 
interested in, is relevant. All the other bound states are supposed to be 
very far away. This allows us to make, in the energy range we are 
interested in, a single pole approximation for $ G _{22} ^{0}$:
\begin{eqnarray}
G _{22} ^{0} ( {\bf  r}, {\bf  r}', \omega ) = \frac{ \varphi ({\bf  r})  
\varphi  ^{*}({\bf  r}')}{ \omega - E _{0} + i \epsilon } 
\label{eq6}
\end{eqnarray}
where $\varphi ({\bf  r})$ is the wavefunction of the bound state and $ 
E _{0} $ its energy. This makes Eq.(5) explicitely soluble because $ G 
_{22} ^{0}$ becomes basically a projector on the bound state. One 
finds:
\begin{eqnarray}
G _{11}= G _{11} ^{0} + \frac{1}{\omega - E _{0} - \delta E _{0}}  
G _{11} ^{0} t | \varphi > < \varphi  | t^{\dagger} G _{11} ^{0}
\label{eq7}
\end{eqnarray}
where $ \delta E _{0} = < \varphi  | t^{\dagger} 
G _{11} ^{0} t | \varphi > $ is a 
complex quantity.

Let us assume for simplicity that there is no background scattering, i.e. 
the long distance potential $ V _{1}$ is zero. In this case $ G _{11} 
^{0}$ is just the free particle propagator. The corresponding T-matrix 
is then given by the last term in Eq.(7) without the $ G _{11} ^{0}$ 
operators. For the scattering we are interested in, we look for matrix 
elements between plane waves with very small wavevectors compared 
to the molecular scale. Since $ \varphi  ( {\bf  r})$ and $ t ( {\bf  r}, 
{\bf  r}')$ are short-ranged we can as well take these wavevectors to be 
zero. This leads to a numerator equal to $|w| ^{2}$ with $ w =  \int d 
{\bf  r} d {\bf  r}' t ( {\bf  r}, {\bf  r}') \varphi ({\bf  r}')$. The 
denominator gives a pole for $ \omega = E$ with $ E = E _{0} + \delta 
E _{0}$, corresponding to the resonance produced by the bound state. 
The real part  Re$E$ gives the physical energy $ \omega _{0}$ of the 
resonance, which is the one actually measured experimentally. So we 
do not have to worry about calculating Re$\delta E _{0}$. The 
imaginary part gives the width of the resonance due physically to the 
possible decay, induced by $t$, of the molecule into two atoms. 
Introducing the Fourier transform $G^{0} _{ {\bf  k}}$ of the free 
particle propagator, this imaginary part will come from Im$G^{0} _{ 
{\bf  k}}= - \pi \delta ( \omega - \epsilon _{ {\bf  k}})$ with $\epsilon 
_{ {\bf  k}}= k ^{2}/2 m _{r}$, physically linked to the density of 
final states for the decay. Since we are concerned with low energy $ 
\omega $, the wavevector must be small and the matrix elements 
coming from $ t | \varphi > $ in the above expression of $ \delta E 
_{0}$ can again be evaluated for zero wavevector, which introduces 
again $|w| ^{2}$. Finally we obtain for the corresponding scattering 
amplitude:
\begin{eqnarray}
f( \kappa ) = - \frac{1}{ (\omega -  \omega _{0})/ \gamma + i \kappa } 
\label{eq8}
\end{eqnarray}
to be evaluated on the shell $ \omega = \kappa ^{2}/ 2 m _{r}$. In this 
expression we have set $ \gamma = m _{r} |w| ^{2}/2 \pi $. We find 
in particular in Eq.(8) that Im$ f ^{-1}( \kappa ) = - \kappa $ as 
required by unitarity. Evaluating Eq.(8) at zero energy gives the 
scattering length $ a = - \gamma / \omega _{0}$. Strictly speaking the 
Feshbach resonance corresponds to the situation where the above 
resonance occurs at zero energy $ \omega  = 0$. This occurs for $ 
\omega _{0} = 0$, i.e. for an infinite scattering length. Experimentally 
$ \omega _{0}$ is controlled by the applied magnetic field. Naturally 
this result for the scattering amplitude is well known \cite{landau}, as 
well as this general way of modeling the Feshbach resonance 
\cite{timm}  as a simple switch between molecular state and diffusion 
states. We have just reformulated this approach in a way which lends 
itself conveniently to generalization in order to include many body 
effects.

We turn now to the case of a dense Fermi gas and assume again for 
simplicity that there is no background scattering. As we have already 
mentionned we take advantage that the gas is dilute on the molecular 
scale to neglect the effect of the other atoms when two atoms are 
scattering due to the Feshbach resonance. In other words we will take 
for the effective interaction the same as the one we had only two atoms 
present, namely $ \Gamma _{00}=  |w| ^{2}/( \omega - E _{0})$ as it 
results from Eq.(5). This is equivalent to retain only ladder diagrams 
for the short range potential. We note that this effective interaction is 
analogous to the one due to phonon exchange in standard 
superconductors, although there are differences. Actually we believe 
that this description should still be correct even if we take into account 
the perturbation due to the other atoms, provided it is not too strong. In 
this case we will still have resonant scattering, with possibly a shift in 
its energy position, a change in its width and its coupling strength. This 
can be accounted for by a modification of the effective parameters of 
the resonance. Moreover we will already find in our description a 
modification of the position and of the width. So the small perturbation 
due to the other atoms will possibly give a small quantitative change, 
but not a qualitative one. Naturally $ \omega $ in our expression for $ 
\Gamma _{00}$ is the energy for the center of mass of the two atoms at 
rest. If their total momentum is $ {\bf  K}$ and their total energy $ 
\omega $ we have to discount the energy associated with the center of 
mass motion and write $ \Gamma _{00}(\Omega)=  |w| ^{2}/( \Omega 
- E _{0})$, with  $  \Omega = \omega + 2 \mu - K ^{2}/4m$. In this 
last expression we have just taken into account that we will take in the 
following the origin of the single particle energy at the chemical 
potential $ \mu $, which produces a change $ 2 \mu $ for the energy of 
two atoms.

We will now explore the simplest consequences of this interaction by 
ignoring fluctuation-like effects and analogous terms, and taking merely 
$ \Gamma _{00}$ as irreducible vertex. With this assumption we can 
write the Bethe-Salpeter equation for the full vertex $ \Gamma ( \omega 
, K)$, which is directly related to the scattering amplitude, as:
\begin{eqnarray}
\Gamma ( \omega , K) =  \Gamma _{00}(\Omega) +  \Gamma 
_{00}(\Omega) \Pi ( \omega , K) \Gamma ( \omega , K)
\label{eq9}
\end{eqnarray}
where $ \Pi ( \omega , K)$ describes the propagation of two atoms and 
is given, in terms of the full thermal propagator of an atom $ G( \omega 
, {\bf  k})$, by:
\begin{eqnarray}
\Pi ( \omega , K) =  - T \sum_{n} \int \frac{ d {\bf k}}{(2\pi )^{3}} 
G( - i \omega - \omega _{n}, {\bf  K}-{\bf  k}) G( \omega _{n} , {\bf  
k})
\label{eq10}
\end{eqnarray}
with $\omega _{n}$ = $ ( 2n + 1 ) \pi T $ being the Matsubara 
frequency. We have written Eq.(9) in a simple way by taking already 
into account that the full vertex $ \Gamma$ depends only, in our case, 
on the total energy $ \omega $ and the total momentum $ {\bf  K}$ of 
the scattering atoms. This results from the fact that the irreducible vertex 
$\Gamma _{00}(\Omega)$ has itself this property. We can more 
simply rewrite Eq.(9) as:
\begin{eqnarray}
\Gamma _{00} ^{-1}(\Omega) =  \Gamma  ^{-1} ( \omega , K) +   \Pi 
( \omega , K)
\label{eq11}
\end{eqnarray}
We will now eliminate the pole location $ E _{0}$ in $\Gamma 
_{00}$, which is not an observable quantity, in favor of the physical 
energy $ \omega _{0}$ of the resonance, which is directly related to the 
scattering length as we have seen. This is done by writing Eq.(11), at 
zero temperature T, for the case of two atoms in vacuum (implying $ 
\mu =0$), at zero energy $ \omega = 0 $ and momentum ${\bf  K}= 
0$. In this case $ G( \omega _{n} , {\bf  k})$ becomes the free 
propagator $ (  i \omega _{n} - \epsilon  _{k} ) ^{-1}$ and the 
frequency summation in $ \Pi $ becomes an easy integration. On the 
other hand for two atoms in vacuum $ \Gamma  ( \omega , K)$ 
becomes $ \Gamma _{0} ( \Omega) \equiv  |w| ^{2}/( \Omega - \omega 
_{0})$, which is essentially the T-matrix. Hence in this particular 
situation Eq.(11) reduces to:
\begin{eqnarray}
- E _{0}/ |w| ^{2} =  - \omega  _{0}/ |w| ^{2} -   \int \frac{ d {\bf 
k}}{(2\pi )^{3}} \frac{1}{2 \epsilon  _{k}} 
\label{eq12}
\end{eqnarray}
Actually this equation is just equivalent to $ \omega _{0}= E _{0} + 
{\rm Re}\delta E _{0}$. Subtracting Eq.(12) from Eq.(11) we obtain:
\begin{eqnarray}
\Gamma  ^{-1} ( \omega , K) =  \Gamma _{0} ^{-1}(\Omega) -   \Pi ( 
\omega , K) -  \int \frac{ d {\bf k}}{(2\pi )^{3}} \frac{1}{2 \epsilon  
_{k}} 
\label{eq13}
\end{eqnarray}
Taken together the last two terms of the r.h.s. of Eq.(13) give an 
integral which converges for large values of $k$, while each term 
separately is divergent. However this divergence is not a real problem. 
It would not be present if we had kept the $k$ dependence of $ < {\bf  
k} | t | \varphi > $ instead of making $ {\bf  k} = 0 $ at the outset. 
Eq.(13) makes clear a general feature, namely the scattering amplitude 
depends not only on the total energy $ \omega $ of the two particles, 
but also on their total momentum. This is due to the term $ \Pi ( \omega 
, K)$ which gives the effect of the other fermions on the scattering 
process. In other words Galilean invariance for this process is 
obviously lost because of the presence of the Fermi sea. It is clear 
physically that the existence of the Fermi sea wil be unimportant when $ 
{\bf  K}$ is very large. On the other hand we expect this feature will be 
most important for $ {\bf  K}=0$ and we will consider only 
this situation in the following.

Quite remarkably the modifications produced by the other fermions are 
already very important when the effect of the interactions is omitted. 
This corresponds to the modification of the scattering due to Pauli 
exclusion. We will restrict ourselves to this particular problem in the 
rest of the paper. In this case we have $ G( \omega _{n} , {\bf  k}) =  (  
i \omega _{n} - \epsilon  _{k} + \mu  ) ^{-1}$ and the calculation of $ 
\Pi ( \omega , K)$ can be reduced to a single integration over the 
momentum $k$. It is convenient to use reduced units to display this 
result. We take $ \mu $ as our energy scale and $ k _{0}$ as a scale for 
wavevector defined by $ \mu = k ^{2}_{0}/ 2m $ (this is the Fermi 
wavevector at $ T = 0 $ and not much different at low temperature). We 
introduce the reduced wavevector $ x = k / k_{0}$, the reduced energy 
$ \bar{ \omega }= \omega / \mu $ and reduced temperature $  \bar{ t }= 
T  / \mu $. Then Eq.(13) becomes:
\begin{eqnarray}
- \frac{2 \pi  ^{2}}{m k _{0}} \frac{1}{ \Gamma ( \omega , 0)} = 
\frac{1}{ \lambda } - \frac{\bar{ \omega }+2 }{ \bar{ W}} + 
\int_{0}^{ \infty}dx [  1 - \frac{x ^{2}}{ x ^{2}-1 - \bar{ \omega }/2}  
\tanh \frac{ x ^{2}-1}{2 \bar{t}}]  
\label{eq14}
\end{eqnarray}
where in the last term $ \bar{ \omega }$ has to be 
understood with an infinitesimal positive imaginary part.
Except for the factor $ \pi /2 k_{0} $ this is just the inverse $f ^{-1}$ 
of the effective scattering amplitude for our problem. We have 
introduced the coupling constant $ \lambda  = 2 k_{0} |a| / \pi  $, also 
related to the detuning $ \omega _{0} $ by 
$ \lambda = ( 2 / \pi ) W/ \omega _{0} $ with $ W = 
\gamma k_{0} $ being the energetic (half) width of the resonance line 
for a wavevector $k_{0}$. We have used the reduced width $  \bar{ 
W} = ( 2 / \pi ) W / \mu $. If we consider the specific case of $ ^{6}$Li 
, the most heavily explored experimentally, the standard energy width 
of the Feshbach resonance is given \cite{timm}  by $ \gamma /  
|a_{bg}| $ where $ a_{bg}$ is the high field limit of the scattering 
length, of order of 100 nm. It is experimentally of order of 100 G, 
which translates into an energy of 10 mK. If we consider dense gases 
for which we have $k_{0} |a_{bg}|  \sim 1$, this gives $W \sim$ 10 
mK. Since we will have at most experimentally $ \mu  \sim $ 10$ \mu 
$K, we see that $  \bar{ W} \sim 10 ^{3}$. Since we are interested in 
reduced energy $ \bar{ \omega }$ of order 1, this makes the second 
term of the r.h.s. completely negligible and we omit it from now on. 
The same is likely to be true in most useful cases of Feshbach 
resonance. Note however that this term is necessary if we want to find 
in the lower complex plane the pole corresponding to the Feshbach resonance.

Let us call $I(\bar{ \omega })$ the integral in Eq.(14). The imaginary 
part of the r.h.s. is easily found to be Im$ I(\bar{ \omega }) = - ( \pi /2 
) R \tanh (\bar{ \omega }/ 4 \bar{t})$ with $ R = ( 1 + \bar{ \omega }/2 
) ^{1/2}$, and is plotted in Fig. 1.  At zero temperature with no Fermi 
sea, this would give us back the imaginary part in Eq.(8). We see that, 
in addition to $ \bar{ \omega }= - 2$ corresponding to zero kinetic 
energy, this imaginary part is zero for $\bar{ \omega }=0$, that is at the 
chemical potential. This is expected on general grounds since injecting 
particles at this energy does not perturb equilibrium and so does not 
lead to decay. More generally the $\tanh (\bar{\omega} / 4 \bar{t})$ can 
be understood as the decrease of the scattering resulting from Pauli 
exclusion on the final state, together with the existence of reverse 
processes, both due to thermal occupation.
\begin{figure}
\vbox to 65mm{\hspace{-6mm} \epsfysize=65mm \epsfbox{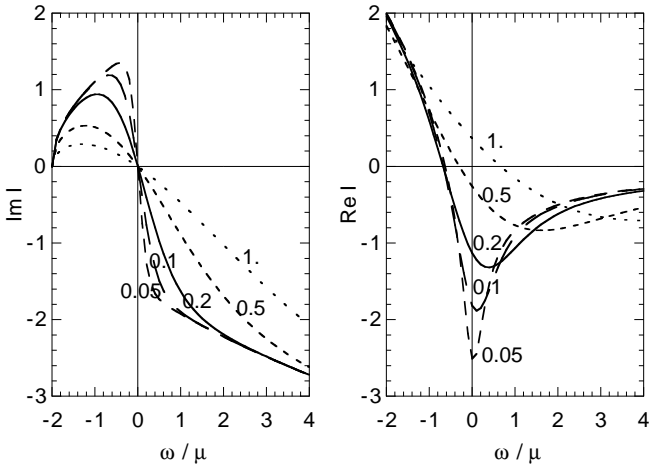} 
}
\caption{Imaginary and real part of the integral $I( \bar{ \omega })$
in Eq.(14), as a function of reduced energy
$ \bar{ \omega }= \omega / \mu $
for various reduced temperature $ \bar{t}=T/ \mu $ indicated
in the figure.}
\label{figure1}
\end{figure}
\vspace{-2mm}
Re$I(\bar{ \omega })$ is plotted in Fig. 1 for various reduced 
temperatures. After adding the term $ 1/ \lambda $ and multiplying by 
the factor $ - \pi /2 k_{0} $, we can consider the result as the inverse of 
an effective scattering length $ a ^{-1} _{eff}$ for two atoms. When $ 
\lambda $ is of order of unity or larger, we see as naturally expected 
that the scale for this scattering length is the only one left in the 
problem, namely $ 1 / k _{0}$. As seen in the figure $ a ^{-1} _{eff}( 
\bar{ \omega })$ has a strong energy dependence on the scale of the 
Fermi energy $E _{F}$. This is in contrast with the case of two 
isolated atoms seen in Eq.(8), where in the same energy range the real 
part of $- f ^{-1}$ is a constant equal to the scattering length $a ^{-1}$ 
(except if $\omega _{0}$ is of order  $E _{F}$ that is extremely near 
the Feshbach resonance). This means that an essential simplification in 
the scattering properties effectively disappears. Indeed for ultracold 
atoms scattering is characterized by a single parameter, namely the 
scattering length $a$. Now because of the Fermi sea the energy scale 
$E _{F}$ appears and the scattering amplitude gets a complex energy 
dependence, which depends also on temperature. An immediate 
consequence is that physically the Feshbach resonance is actually 
washed out by the Fermi sea. Indeed instead of having for all possible 
scattering atoms a diverging scattering length, and correspondingly a 
zero Re$ f ^{-1}$, we have a Re$ f ^{-1}$ which is of order $ 1 / k 
_{F}$ and depends on the energy of the two considered atoms (as well 
as their momentum as we have seen). This occurs as soon as $ \lambda 
$ is not small. In particular nothing special occurs right at the Feshbach 
resonance when  $ \lambda ^{-1}=0 $. This provides a simple 
explanation to the experimental observations \cite{thomas} that the 
resonance is not seen when the magnetic field is swept through its 
assumed location when the gas is dense enough to be in the degenerate 
regime.
Naturally the inhomogeneity due to the varying trapping potential is an 
additional source of smearing since the energy scale $ \mu $ for the 
scattering is space dependent.

We look now more strictly for a resonance where the scattering 
amplitude diverges, which for two isolated atoms occurs at zero energy 
at the Feshbach resonance $ a ^{-1}=0$. So we require that both the 
real and the imaginary part of $ \Gamma ( \omega , 0) ^{-1}$ are zero 
in Eq.(14). It is easy to see that, for $ \bar{ \omega }= - 2$, the real 
part of the integral in Eq.(14) is always positive so Re$ f ^{-1}>0$ for 
$a<0$. So the only possible resonance occurs at the chemical potential 
$\bar{ \omega }=0$. In this case the condition that Re$ f ^{-1}=0$ in 
Eq.(14) coincides with the well-known condition for the BCS pairing 
instability. In particular, at $T=0$, the logarithmic divergence of Re$ f 
^{-1}$ which occurs for $\bar{ \omega }=0$ is a mark of this 
instability. It is known that this pairing instability is basically due to 
Pauli exclusion by the Fermi sea on low energy states, which produces 
a shift from a 3D situation to an effective 2D physics. We can 
understand qualitatively the strong energy dependence of the scattering 
amplitude we have found above as a manifestation of this 2D physics. 
Naturally, since we have not included interactions, the value of the 
critical temperature $T _{c}$ we have for the pairing instability is just 
the standard one \cite{stoofal}, and it does not contain lower order 
fluctuation effects \cite{gork} nor higher orders and self-energy effects 
considered in recent calculations \cite{rc}. Obviously our calculation is 
no longer strictly valid below $T _{c}$ since we should deal with the 
superfluid state, which can be done as we have already mentionned. 
However this is not an important restriction for our present purpose 
since we want to stress the strong frequency dependence of Re$ f ^{-
1}$ on the scale of $E _{F}$, which will clearly remain at least 
qualitatively valid in the superfluid state. In the same spirit we have not 
included interactions, but they will not change the basic Pauli exclusion 
physics. We can not expect interactions to remove the energy 
dependence of the scattering and our qualitative conclusion will remain 
unchanged.

In conclusion we have presented a coherent framework which allows to 
deal with many-body effects in the presence of a Feshbach resonance. 
As a simple consequence we have shown that the mere result of Pauli 
exclusion, which results from Fermi statistics, induces a strong 
modification of the scattering properties. It is clear that this modification 
is a necessary ingredient in the physical understanding of these systems 
since Pauli exclusion can not be ignored. This modification results in a 
smearing of the Feshbach resonance and provides a natural explanation 
for recent experimental findings.

We are very grateful to Y. Castin, C. Cohen-Tannoudji, J. Dalibard, X. 
Leyronas, C. Mora and C. Salomon for very stimulating discussions.

* Laboratoire associ\'e au Centre National
de la Recherche Scientifique et aux Universit\'es Paris 6 et Paris 7.


\begin{references}
\bibitem{jin}B. DeMarco and D. S. Jin, Science {\bf  285}, 1703 
(1999); A. G. Truscott \emph{et al}, Science {\bf  291}, 2570 (2001); 
F. Schreck \emph{et al}, Phys.Rev.Lett. {\bf 87}, 80403  (2001).
\bibitem{thomas}K. Dieckmann  \emph{et al}, Phys.Rev.Lett. {\bf 
89}, 203201  (2002); K. M. OÕHara \emph{et al}, Science {\bf  298}, 
2179 (2002); J. Cubizolles \emph{et al}, preprint.
\bibitem{stoofal}  H. T. C. Stoof, M. Houbiers, C. A. Sackett and R. 
G.Hulet, Phys. Rev. Lett. {\bf 76}, 10 (1996).
\bibitem{timm}E. Timmermans, P. Tommasini, M. Hussein and A. 
Kerman, Phys.Reports {\bf  315}, 199 (1999) and references therein.
\bibitem{holland}M. Holland \emph{et al}, Phys.Rev.Lett. {\bf 87}, 
120406  (2001).
\bibitem{landau}L. D. Landau and E. M. Lifshitz, \emph{Quantum 
Mechanics} (Pergamon Press, Oxford, 1980).
\bibitem{bardeen}J. Bardeen, Phys.Rev.Lett. {\bf 6}, 57  (1961).
\bibitem{caroli}C. Caroli \emph{et al}, J. Phys. C {\bf 5}, 21 (1972).
\bibitem{gork}  L. P. Gorkov and T. K. Melik-Barkhudarov, Sov. 
Phys. JETP {\bf 13}, 1018 (1961).
\bibitem{rc}R. Combescot, Phys.Rev.Lett.{\bf 83}, 3766 (1999).

\end{references}
\end{document}